\documentclass[conference]{IEEEtran}
\IEEEoverridecommandlockouts
\usepackage{cite}
\usepackage{amsmath,amssymb,amsfonts}
\usepackage{algorithmic}
\usepackage{graphicx}
\usepackage{textcomp}
\usepackage{xcolor}
\usepackage{float}
\usepackage{graphicx}
\usepackage{bbding}
\usepackage{array}
\usepackage{enumerate}
\def\BibTeX{{\rm B\kern-.05em{\sc i\kern-.025em b}\kern-.08em
		T\kern-.1667em\lower.7ex\hbox{E}\kern-.125emX}}

\begin{document}

\title{A privacy-preserving, decentralized and functional Bitcoin e-voting protocol\\
{\footnotesize}
\thanks{Acknowledgement. The authors thank the editors and the anonymous reviewers for their valuable comments. This study is supported by the National Science foundation of China (No. 61472074,U1708262) and the Fundamental Research Funds for the Central Universities (No.N172304023).}
}

\author{\IEEEauthorblockN{1\textsuperscript{st} Zijian Bao}
\IEEEauthorblockA{\textit{Department of Computer Science and Engineering} \\
\textit{Northeastern University}\\
Shenyang, China \\
zijianbao@gmail.com}
\and
\IEEEauthorblockN{2\textsuperscript{nd} Bin Wang}
\IEEEauthorblockA{\textit{Department of Computer Science and Engineering} \\
\textit{Northeastern University}\\
Shenyang, China \\
binge1638@163.com}
\and
\IEEEauthorblockN{3\textsuperscript{rd} Wenbo Shi}
\IEEEauthorblockA{\textit{Department of Computer Science and Engineering} \\
\textit{Northeastern University}\\
Qinhuangdao, China \\
swb319@hotmail.com}
}

\maketitle

\begin{abstract}
Bitcoin, as a decentralized digital currency, has caused extensive research interest. There are many studies based on related protocols on Bitcoin, Bitcoin-based voting protocols also received attention in related literature.

In this paper, we propose a Bitcoin-based decentralized privacy-preserving voting mechanism. It is assumed that there are n voters and m candidates. The candidate who obtains t ballots can get x Bitcoins from each voter, namely nx Bitcoins in total. We use a shuffling mechanism to protect voter's voting privacy, at the same time, decentralized threshold signatures were used to guarantee security and assign voting rights. The protocol can achieve correctness, decentralization and privacy-preservings. By contrast with other schemes, our protocol has a smaller number of transactions and can achieve a more functional voting method.
\end{abstract}

\begin{IEEEkeywords}
Bitcoin; Anonymous voting; shuffling mechanism
\end{IEEEkeywords}

\section{introduction}
Voting plays an important role in modern life. Electronic voting has aroused the attention of many scholars for a long time. However, how to design a voting protocol which is decentered and privacy-preserving is an open issue. Bitcoin\cite{nakamoto2008bitcoin}, as a new type of decentralized digital currency, has a wide range of applications in the fields of voting\cite{tian2017simpler,zhao2015vote,bartolucci2018sharvot}, secure multiparty computations\cite{andrychowicz2014secure}, public randomness source\cite{bonneau2015bitcoin} and designing fair protocols\cite{bentov2014use}. 

In the field of voting, Zhao and Chan\cite{zhao2015vote} first proposed how to vote privately using Bitcoin. There are n voters that each has 1 Bitcoin and votes for 2 candidates. The winner can obtain all the n Bitcoins. The scheme only supports $1\text{-of-}2$ election mode. Tian et al.\cite{tian2017simpler} propose a simple Bitcoin voting protocol which can produce a ballot by a voter selecting at least ${{\text{k}}_{\min }}$ at most ${{\text{k}}_{\max }}$ winners from L candidates. Silvia et al.\cite{bartolucci2018sharvot} proposed the circle shuffle mechanism for Bitcoin voting to provide privacy protection, but it requires a centralized dealer. Meanwhile, there are many papers focusing on solving the 
Bitcoin anonymity problem\cite{conti2017survey}, coinjoin\cite{maxwell2013coinjoin} was first to achieve security against stealing mixes by using group transactions. However, it requires a centralized service to confuse output addresses. CoinShuffle\cite{maxwell2013coinjoin} improves over Coinjoin by using decryption mixnets for address shuffling which achieves anonymity against insiders. It has a flaw that the last one of the shuffling may put his own output address in the specified location.

The main goals of bitcoin-based e-voting protocols should include:
\begin{itemize}
\item Correctness: The most basic and important nature of an e-voting agreement is to ensure the correctness of voting which prevents voters from being falsified, discard or repeat votes.
\item Decentralization: In the entire voting process, in addition to voters and candidates, no other third-party institutions or trusted agencies are required to assist in the whole process.
\item Privacy protection: Voting information of voters cannot be known by anyone else. In reality, privacy protection is one of the most important attributes of voting protocol.
\item Functionality: More forms of voting should be supported, such as office voting, large-scale election voting, $1\text{-of-}2$  candidate voting, multiple candidate voting, etc.
\end{itemize}

\begin{figure*}[htb]
\centerline{\includegraphics[width=7.5in]{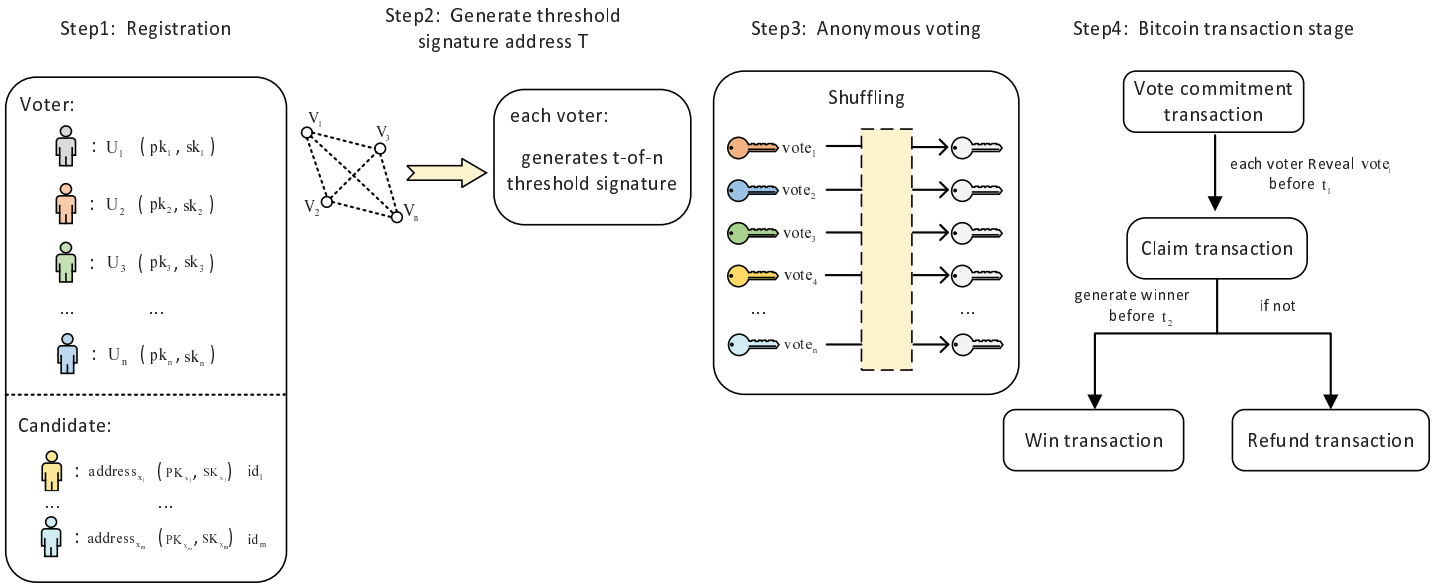}}
\caption{Entire voting protocol flow diagram.}
\label{Fig.1}
\end{figure*}

\section{preliminaries}

\subsection{Bitcoin transactions}

In this article, we do not consider attacks on Bitcoin such as 51\% attacks, routing attacks\cite{conti2017survey}, as well as transaction fee. We assume that all voters and candidates have access to the Bitcoin network and the blockchain does not will be forked.
  
In a simplified bitcoin model, Bitcoin transaction contains inputs, outputs and value prices. Output can be seen as a validation of transaction and input script is the papameters(e.g. signature of the previous input)for the program script in the output. Among them, optional item $\text{locktime}\left( \text{t} \right)$\cite{andrychowicz2014secure} can ensure that the transaction will become efftive only after a period of time t. After(t)\cite{todd2014bip} means that a single output script can be made unspendable until t time.

This article also uses a special form of Bitcoin script, P2SH\cite{antonopoulos2014mastering} (Pay-to-Script-Hash), which was introduced in 2012 as a new, powerful transaction type that greatly simplifies complex scripts. We use a specific script, the output script is:
\[\text{OP}\_\text{HASH}160<\text{Hash}\left( \text{x} \right)>\text{OP}\_\text{EQUAL}.\]
The input script is x. As to get the corresponding Bitcoins, user needs to expose the value of x.

\subsection{Decentralized threshold signature}\label{threshold}

Threshold secret sharing\cite{shamir1979share,goldfeder2015securing} is a way to split a secret value into several shares that can be given to different participants. However, there is an important issue is how to generate and distribute these shares. The simplest way is to introduce a trusted dealer who begins with the constructed key, generates the shares and distributes them to each party. Of course, this has a weakness in that the trusted dealer is a single point of failure. Another way is to generate shares of a key in a distributed manner without ever constructing the key in the process. Scheme\cite{goldfeder2015securing} is the ECDSA scheme that works for arbitrary n and any ${\text{t}<\text{n}}$ which is also compatible with Bitcoin. 

While most Bitcoin transactions are spent with a single signature, Bitcoin in fact specifies a script written in a stack-based programming language which defines the conditions under which a transaction may be redeemed. This scripting languag  require at least $\text{t-of-n}$ specified ECDSA public keys to provide a signature on the redeeming transaction. A relatively recent feature of Bitcoin, pay-to-scripthash,  enables payment to an address that is the hash of a  script. When this is used, senders specify a script hash,  and the exact script is provided by the recipient when  funds are redeemed.  A quirk of pay-to-script hash is that the  ${\text{n}\leq\text{3}}$ restriction is removed from t-outof-n multisignature transactions. However, due to a hardcoded limit on the overall size of a hashed script, the recipients are still limited to ${\text{n}\leq\text{15}}$.

All of our constructions that use threshold signatures can be instantiated with the threshold signature scheme in \cite{goldfeder2015securing}.
We argue that threshold signatures offer fundamental advantages stemming from the fact that in the multisignature approach:
\begin{itemize}
\item Flexibility. Threshold signatures are more flexible than multisignatures in the access policies that they permit as well as in the ability to modify the access policies. Threshold signatures also allow more flexibility for making changes to the access control policy.

\item Anonymity. While Bitcoin allows users to be pseudonymous, it does not provide any anonymity guarantees. Indeed, it has been shown that it is not difficult to link various addresses belonging to a single user. Moreover, because the entire transaction log is public, once an address has been associated with a real world identity, one can immediately view every other transaction associated with that address.

\end{itemize}
\section{our voting protocol}

The entire process of our scheme is shown in Fig.\ref{Fig.1}. The specific process includes registration on the bulletin board\cite{rafaeli1984electronic}, generation of the threshold signature address, anonymous voting, and Bitcoin transaction stage. We assume that our Bitcoin voting protocol is used in a small-scale, limited-power scenario. Suppose there are n voters, each one has his own Bitcoin address and sufficient balance, and there are m candidates, if one of them gets t votes or more, then he can get x Bitcoins from each voter.

\subsection{Registration}\label{AA}
For a voter ${{\text{V}}_{\text{i}}}$, each voter needs to have an $\text{address i}$ (abbreviated as ${{\text{U}}_{\text{i}}}$) which contains at least x + z Bitcoins that x represents the Bitcoins used to vote and z is used to guarantee the security of the decryption of vote commitment, the voter needs to generate and publish his own key pair $\text{(p}{{\text{k}}_{\text{i}}}\text{, s}{{\text{k}}_{\text{i}}})$ for the shuffling operation in Section \ref{section B}. All voters negotiate the last time ${{\text{t}}_{\text{1}}}$ of revealing the vote commitment and the latest time ${{\text{t}}_{2}}$ for returning the deposit. If there is no candidate to win finally. Each candidate needs to prepare an address to obtain Bitcoin after winning and reveal the key pair $\text{(P}{{\text{K}}_{\text{x}}},\text{S}{{\text{K}}_{\text{x}}})$ (x indicates the id of the candidate, $\text{x}\in \left\{ 1,2,\cdots ,\text{m} \right\}$). All the above information is published on the bulletin board. Related information is also shown in Fig.\ref{Fig.1}.

\subsection{Generation of the threshold signature address}\label{section B}
We use the decentralized threshold signature scheme mentioned in Section\ref{threshold} and each voter interacts to generate the $\text{t-of-n}$ threshold signature address $\text{T}$. For each voter's $\text{s}{{\text{k}}_{{{\text{T}}_{\text{i}}}}}$ which is the share of $\text{T}$ is considered as a valid vote. Each voter can vote his share to the candidate whom he supports and the candidate who reaches $\text{t}$ shares can receive the  nx Bitcoin rewards at last.

\subsection{Anonymous voting}\label{shuffling}
\begin{figure}[htb]
\centerline{\includegraphics[width=3.5in]{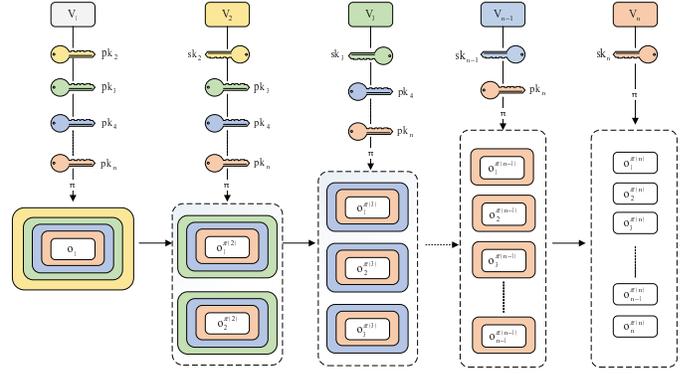}}
\caption{Anonymous voting based on shuffling.}
\label{Fig.2}
\end{figure}

Based on the relevant design of the shuffle\cite{ruffing2014coinshuffle}, we propose a new anonymous voting mechanism. Also, we solved the defect for the last one of shuffling to place his own vote in a specific position and provided additional verification. This phase is illustrated in Fig.\ref{Fig.2}.
\begin{enumerate}[i.]
\item Voter ${{\text{V}}_{1}}$ has already known the remaining voters' public key $\text{p}{{\text{k}}_{\text{i}}}$ through the bulletin board and generates his own vote according to his selection. $\text{vot}{{\text{e}}_{\text{1}}}\text{ = }{{\text{E}}_{\text{P}{{\text{K}}_{\text{x}}}}}\text{(s}{{\text{k}}_{{{\text{T}}_{1}}}}||\text{i}{{\text{d}}_{\text{x}}})$, x indicates the candidate chosen by the voter. The $\text{vot}{{\text{e}}_{1}}$ is encrypted by using $\text{p}{{\text{k}}_{2}},\text{p}{{\text{k}}_{3}},\cdots ,\text{p}{{\text{k}}_{\text{n}}}$ to generate $\text{O}_{1}^{\pi (1)}\text{= }{{\text{E}}_{\text{p}{{\text{k}}_{2}}}}\text{(}{{\text{E}}_{\text{p}{{\text{k}}_{3}}}}\text{(}\cdots {{\text{E}}_{\text{p}{{\text{k}}_{\text{n}}}}}(\text{vot}{{\text{e}}_{1}})))$, where $\pi \left( 1 \right)$ is a random permutation offered by ${{\text{V}}_{1}}$. Constructing set ${{\text{O}}_{1}}\text{=}\left\{ \text{O}_{1}^{\pi (1)} \right\}$ and sending it to ${{\text{V}}_{2}}$.

\item ${{\text{V}}_{2}}$ gets $\text{O}_{1}^{\pi (2)}\text{= (}{{\text{E}}_{\text{p}{{\text{k}}_{3}}}}\text{(}\cdots {{\text{E}}_{\text{p}{{\text{k}}_{\text{n}}}}}(\text{vot}{{\text{e}}_{1}})))$ after decryption using his own private key $\text{s}{{\text{k}}_{2}}$. Meanwhile, ${{\text{V}}_{2}}$ selects his $\text{vot}{{\text{e}}_{\text{2}}}\text{ = }{{\text{E}}_{\text{P}{{\text{K}}_{\text{x}}}}}\text{(s}{{\text{k}}_{{{\text{P}}_{2}}}}||\text{i}{{\text{d}}_{\text{x}}})$ and generates $\text{O}_{2}^{\pi (2)}\text{= (}{{\text{E}}_{\text{p}{{\text{k}}_{3}}}}\text{(}\cdots {{\text{E}}_{\text{p}{{\text{k}}_{\text{n}}}}}(\text{vot}{{\text{e}}_{2}})))$ and then sends ${{\text{O}}_{2}}\text{=}\left\{ \text{O}_{1}^{\pi (2)}\text{,O}_{2}^{\pi (2)} \right\}$ to ${{\text{V}}_{3}}$ after construction.

\item The rest can be done in the same manner, until ${{\text{V}}_{\text{n}}}$ gets the final ${{\text{O}}_{\text{n}}}\text{=}\left\{ \text{O}_{1}^{\pi (\text{n})}\text{,O}_{2}^{\pi (\text{n})}\cdots \text{O}_{\text{n}}^{\pi (\text{n})} \right\}\text{=}\left\{ \text{vot}{{\text{e}}_{1}},\text{vot}{{\text{e}}_{2}},\cdots ,\text{vot}{{\text{e}}_{\text{n}}} \right\}$, the lexicographic order is then sent to all voters.

\item Each voter hashes the content to get $\text{H}\left( \text{i} \right)$ and broadcasts to each other to determine if all $\text{H}\left( \text{i} \right)$ are equal. If equals, each one generates the last round of random permutation $\pi \left( \text{n}+1 \right)$ with pseudo-random number generator\cite{phillips2011pseudo} and $\text{H}\left( \text{i} \right)$. The purpose of this round is mainly to prevent ${{\text{V}}_{\text{n}}}$ from being able to place his vote in specific positions.
\end{enumerate}

\subsection{Bitcoin transaction stage}
\begin{figure*}[htb]
\centerline{\includegraphics[width=7in]{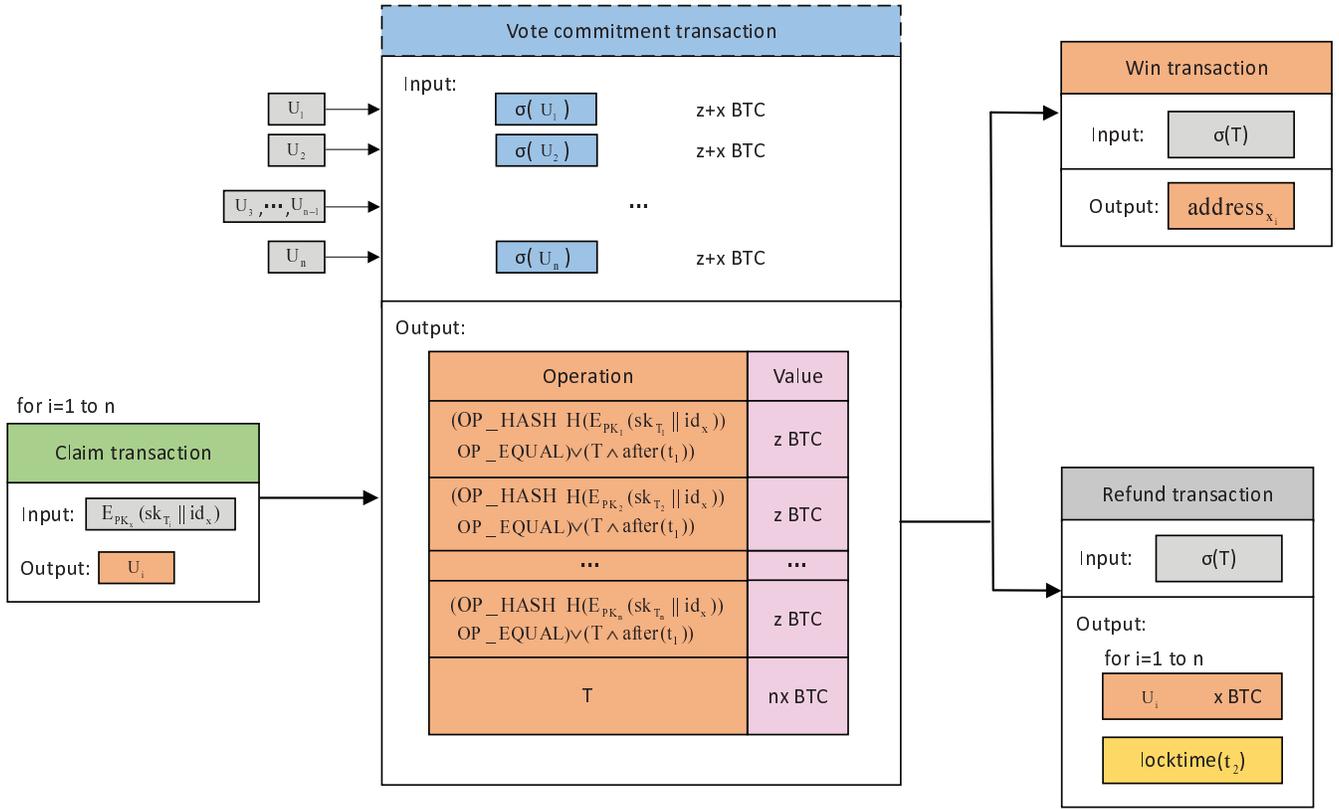}}
\caption{Bitcoin transactions stage.}
\label{fig3}
\end{figure*}

After completing the shuffle operation, any voter can construct vote commitment transaction and refund transaction (see Fig.\ref{fig3}) and then send them to all voters for signature. When the signature is completed, one voter just publishes the vote commitment transaction to Bitcoin network and the refund transaction is kept locally until time ${{\text{t}}_{2}}$ which is the last time for returning ballot funds.
The input of vote commitment transaction includes the address $\text{i}$ which owns $\text{z+x}$ Bitcoins, $\text{i}\in \{1,2,\cdots ,\text{n}\}$.

The output includes two aspects:
\begin{itemize}
\item
There are 2 ways to take away z Bitcoins.
\[
(\text{OP}\_\text{HASH}160<\text{Hash}({{\text{E}}_{\text{P}{{\text{K}}_{\text{x}}}}}\text{(s}{{\text{k}}_{{{\text{T}}_{i}}}}||\text{i}{{\text{d}}_{\text{x}}}))>\text{OP}\_\text{EQUAL})
\]
Firstly, it means voter can reveal ${{\text{E}}_{\text{P}{{\text{K}}_{\text{x}}}}}\text{(s}{{\text{k}}_{{{\text{T}}_{1}}}}||\text{i}{{\text{d}}_{\text{x}}})$ to obtain his own deposit z Bitcoins. 
\[(\text{T}\wedge \text{After}({{\text{t}}_{1}}))\]
Secondly, it means that the deposit can be taken away by the voters jointly construct the transaction or by the winner candidate who obtians the private key of ${\text{T}}$. The purpose of setting a deposit is to prevent voter from refusing to vote. Therefore, the handle of z Bitcoin deposit can actually be based on specific actual demands and designs.

\item T is the address which owns $\text{nx}$ Bitcoins. For any candidate, when the received number of $\text{s}{{\text{k}}_{{{\text{T}}_{\text{i}}}}}$ is greater than $\text{t}$, he can construct a win transaction to obtain the voting reward.
\end{itemize}

Claim transaction. Each voter reveals his $\text{vot}{{\text{e}}_{\text{i}}}$ before time ${{\text{t}}_{1}}$. They revealed ${{\text{E}}_{\text{P}{{\text{K}}_{\text{x}}}}}\text{(s}{{\text{k}}_{{{\text{T}}_{\text{i}}}}}||\text{  i}{{\text{d}}_{\text{x}}})$ to recover the deposit simultaneously, at the same time, the candidate corresponding to x decrypt $\text{sk}_{{{\text{T}}_{\text{i}}}}||\text{  i}{{\text{d}}_{\text{x}}}$ with his private key $\text{S}{{\text{K}}_{\text{x}}}$ on the Bitcoin network.

Win transaction. Once a candidate has collected t different $\text{s}{{\text{k}}_{{{\text{T}}_{\text{i}}}}}$, he can initiate a win transaction and transfer nx Bitcions to his own $\text{addres}{{\text{s}}_{\text{x}}}$ to win the vote.

Refund transaction. If after time ${{\text{t}}_{2}}$, the voting result fails to be generated which means no candidate has obtianed enough $\text{t}$ shares, the refund transaction will be triggered, then each voter's $\text{x}$ Bitcoins will be returned to the original address ${{\text{U}}_{\text{i}}}$.

\section{discussion}

First of all, our solution is correct, vote commitment transaction can ensure that each voter has only one vote. They cannot tamper, discard or repeat the vote after being sent to Bitcoin network, discard and repeat the voting. The protocol is decentralized, because of the use of a non-central threshold signature algorithm, the decentralized shuffling mechanism and Bitcoin is also decentralized. Privacy protection is mainly in the shuffling mechanism in Section \ref{shuffling}. It uses the decryption mixnets to achieve the purpose of protecting vote privacy. The specific security analysis can be seen in \cite{ruffing2014coinshuffle}. Considering the number of transactions, our protocol needs a vote commitment transaction, n claim transactions, and win/refund transaction in the Bitcoin network, so the complete process requires $\text{n+}2$ transactions. For functionality, we can complete the $\text{t-of-n}$ voting form. For example, we hope to hold a election that winner should have more than $50\%$ votes. We can set $\text{n}$ as the number of voters and $\text{t}$ as half the number of voters. Similarly $2/3$,       $4/5$ etc. It makes voting forms actually more flexible.

\renewcommand\arraystretch{2}
\begin{table}[]
\centering
\caption{comparsion between our protocol with others.}
\label{my-label}
\begin{tabular}{p{1.65cm}<{\centering}|p{1.5cm}<{\centering}|p{1.5cm}<{\centering}|p{1.55cm}<{\centering}|p{1.1cm}<{\centering}}
\hline
& \begin{tabular}[c]{@{}l@{}}Zhao et al\cite{zhao2015vote}\end{tabular} & \begin{tabular}[c]{@{}l@{}}Tian et al\cite{tian2017simpler}\end{tabular} & \begin{tabular}[c]{@{}l@{}}Silvia et al\cite{bartolucci2018sharvot}\end{tabular} & \begin{tabular}[c]{@{}l@{}}Our\end{tabular}  \\ \hline
Correctness            &\Checkmark           &\Checkmark                  &\Checkmark                 & \Checkmark \\ \hline
Decentralization     &\Checkmark           &\Checkmark                  &\XSolid                          & \Checkmark \\ \hline
Privacy protection     &\Checkmark          &\Checkmark                  &\Checkmark                  & \Checkmark \\ \hline
Transaction numbers & $2\text{n+}2$              &$\text{n+}2$                               &$\text{n+1}$                                &$\text{n+}2$ \\ \hline
Functionality          & $1\text{-of-}2$                      &$\left[ \text{k}{}_{\min },{{\text{k}}_{\max }} \right]$ winners from L candidates          &$\text{t-of-n}$ $\forall\text{t},\text{t}\leq\text{n}$  &$\text{t-of-n}$ $ \forall\text{n},\text{t}\leq\text{n}$\\ \hline
\end{tabular}
\end{table}

Zhao and Chan's solution\cite{zhao2015vote} used an complex cryptography tool such as zero-knowledge proof, it needs to run the zkSNARK 3 times, $\text{n-}1$ secure unicast and 2 times public broadcast, and the number of transactions is $2\text{n+}2$, the scheme can only achieve 1-of-2 election form. Tian et al.\cite{tian2017simpler} proposed improvements based on the Chan's protocol that can reduce the number of transactions to $\text{n+}2$. Unfortunately, it also needs $\text{n}\left( 2\text{L+}1 \right)$ proofs of zkSNARK which L is the number of candidates, $\text{L}\ge 1$. Silvia et al.\cite{bartolucci2018sharvot} put forward the circle shuffle technique for Bitcoin voting, which also requires only $\text{n+}1$ transactions, but it requires a centralized honest and trustable dealer. Once the dealer is malicious, the entire vote protocol will be destroyed. Table \ref{my-label} summarizes the main differences of our protocol with others.

Future work. Our voting protocol is set in a small-scale (number of voters restrictions), permission voting scene. For large-scale voting, we recommend the use of centralized shamir's secret sharing program\cite{shamir1979share}, but in fact, Bitcoin is not a dedicated voting system, its performance will be limited. Designing a blockchain system that fully serves voting will probably solve this problem better.

\section{conclusion}

We proposed a privacy-preserving, decentralized and functional Bitcoin e-voting protocol that uses a shuffling mechanism to complete privacy protection, decentralized threshold signatures to assign voting rights, use of Bitcoin transaction to make all voting transparency and immutability, and the P2SH scripts can prevent the phenomenon of discarding vote. The protocol reaches the correctness, decentralization, privacy protection and has more flexible voting forms. Meanwhile, the number of transactions maintaines small.

\vspace{12pt}

\end{document}